# A Novel Agent-Based Simulation Framework for Sensing in Complex Adaptive Environments

Muaz A. Niazi, *Member IEEE*, and Amir Hussain, *Senior Member IEEE*

*Abstract*—In this paper we present a novel Formal Agent-Based Simulation framework (FABS). FABS uses formal specification as a means of clear description of wireless sensor networks (WSN) sensing a Complex Adaptive Environment. This specification model is then used to develop an agent-based model of both the wireless sensor network as well as the environment. As proof of concept, we demonstrate the application of FABS to a boids model of self-organized flocking of animals monitored by a random deployment of proximity sensors.

## I. INTRODUCTION

WIRELESS Sensor Networks (WSN) are a relatively recent development in the domain of distributed systems. A WSN comprises a set of small computing devices capable of sensing and reporting various parameters from their environment. WSNs and their counterparts, Sensor Actuator Networks (SAN) are used to remotely sense and, at times, act on various parameters in almost any conceivable type of environment. Being a relatively new technology, WSNs have been used primarily to sense simple parameters such as temperature, humidity etc. Gradually however, WSNs are finding applications in an increasing set of complex sensing problems. Although there has been considerable interest in this domain, certain complexities associated with designing an effective WSN based application make this task non-trivial. These problems are primarily of two types; firstly sensors are occasionally deployed on a one-time basis, and secondly, as we shall examine further in this paper, sensors might need to perform cognitive tasks by being able to intelligently sense complex phenomenon in the environment. To tackle these problems, new sensor paradigms such as Cognitive Sensor Networks[1] are being proposed to develop advanced sensing applications.

Although sensors are gradually becoming smaller and less costly, most Wireless Sensor Network applications currently rely on Modeling and Simulation (M&S) as proof of concept. Thus, being able to develop a realistic and credible simulation model is an effective M&S strategy which is imperative to the design of a successful WSN application.

Up until recently, M&S has mainly been used for simulation of basic communication paradigms primarily due to the uncertainty and limitations associated with sensor networks. These include simulations related to data retrieval, fusion, aggregation and basic communication issues (such as in the MAC layer). As such, there has been a lack of focus on realistic modeling of the dynamism associated with sensing real world Complex Adaptive environments. However, it is our belief that, with the rapid growth of Wireless Sensor Network applications, there is now a growing need for a renewed focus on the realistic modeling of the sensed environment. This need which also requires shifting of focus from lower to higher layers using abstraction is akin to a segregation of layers in wired networks where applications can be developed without worrying about the workings of the lower layers.

In general, realistic simulations of Wireless Sensor Networks (WSNs) [2] can be a challenge to develop due to several reasons. Firstly, everything related to WSNs is about change. Secondly, design of a WSN application of monitoring an even moderately complex phenomenon can be a challenge because an environment can consist of any number and combination of both static and mobile objects, ranging from both organic (living) and inorganic domains. With the wide variety of emerging applications of WSNs, it is becoming increasingly obvious that this lack of available M&S solutions for WSNs situated in Complex environment can severely limit the possible application design using WSN simulations. Thus, any simulation paradigm, which does not also realistically model the environment, would be a weaker representation of the actual system. Other weaknesses of current simulators have been noted in [3].

Environments can range anywhere from the static and immobile such as terrain (mountains, rocks, walls, water etc) having possible impact on communication, placement and routing of the WSN to dynamic (including temporal such as gradual or abrupt changes in the tree cover etc.). At times, the environment can even consist of mobile entities such as animal or insect movements [4].

Most current work on modeling of WSNs [5] uses either

Manuscript received March 20, 2010.
M. A. Niazi is a Doctoral researcher at the Department of Computing Science and Mathematics, University Of Stirling, Scotland and a faculty member at the Department of Biosciences, COMSATS institute of IT, Islamabad. (e-mail: man@ cs.stir.ac.uk)
A. Hussain is with the Department Of Computing Science and Mathematics, University of Stirling, Scotland, UK (e-mail: ahu@cs.stir.ac.uk).

Monte Carlo modeling [6, 7] or differential equation based approaches [8, 9]. However, neither of these effectively solve the problem of modeling the interaction of the WSN with the environment. So, where differential equation models typically focus on the entire system approach, modeling only the aggregate behavior, Monte Carlo methods can be used to focus on individual sensors. Some limitations of Monte Carlo methods have been pointed out in the literature [10].

In this paper, we address this problem of modeling a Complex environment by presenting a novel Formal Agent Based Simulation framework (FABS). FABS can be useful for developing agent-based simulations of wireless sensor networks along with complex adaptive environments. It consists of the following parts:

1. A formal specification based on an ISO standard formal specification language "Z" [11] to unambiguously represent both the sensor network as well as the Complex Adaptive System environment.
2. An agent-based model of a sensor network sensing a self-organized "flocking" environment based on the original "Z" model. This agent-based model extends from the formally developed model in the first step.

The rest of the paper is structured as follows: We first give a brief background required for an understanding of the proposed methodology. This includes an overview of Agent-Based Modeling, basics of formal specification as well as a description of the classical "Boids" model of flocking animals. Next, we develop a formal model of the Complex Adaptive System (CAS) problem composed of a sensor network and a mobile flock of self-organizing "boids" which uses nature-inspired spatial algorithms to perform flocking. Subsequently, we present the agent-based model developed from this formal specification, followed by demonstrative simulation results. Finally, following a brief discussion of the framework's utility and a review of related works, we present some concluding remarks and future work proposals.

## II. BACKGROUND

### A. Agent-Based Modeling

Originating from the domain of social simulation[12], Agent-based modeling (ABM) has been used to model complex problems in domains as diverse as Economics[13], Biology [14], social systems[15], ambient intelligent, peer-to-peer pervasive systems [16], complex computer networks [17] and management of wireless sensor networks[18].

The real strength of Agent-Based Models lies in the representation of Complex Adaptive Systems. In a CAS, the local interactions of the agents result in global phenomena such as emergence and attractors. Examples of emergence include swarming in insects, informal consensus formation in crowds, traffic jams in transportation systems, unplanned clapping and standing ovations in speeches and sports etc. [19]) as well as flocking in animal life forms (e.g. Birds, fish, insects etc. [20]).

### B. Formal Specification

Formal methods in general and formal specification, in particular, is a mechanism employed for mathematical modeling of various types of systems [21]. Using formal specification, the system designer can ensure that the developed model is complete in all aspects required at the selected level of abstraction. Bowen [22] notes that if formal notation is used early during the design of systems, it helps eliminating future errors. One particular aspect of formal specification which makes it particularly effective in modeling real-world phenomena is its representational ability for modeling of concepts at various levels of abstraction. At times, formal methods have also been used to automatically prove theorems about the correctness of a system. "Z" is a formal specification language developed in the 1970's at Oxford University and has since been made an ISO standard. Z is very close to being an executable specification [23] since it can be easily converted to computer code. A Z specification consists of Mathematical notation containing both set-like and first-order predicate calculus based notations. It is written to appear as a group of either state or operation "schemas". Z has been used to model real-world high performance systems such as IBM CICS [24] as well as for modeling Complex reasoning and logics related to the domain of agents and Multi-agent Systems [25].

Although formal specification has been used in other domains, it has never previously been used for the development of Agent-Based Models of Complex Adaptive Systems and Sensor Networks. Our justification of using Z for modeling actually follows from the observation that the current modeling methodologies for WSNs do not provide any guarantees and thus miss out important elements. Specifically, a case in point could be the missed specifications relating to the modeling of Complex Adaptive environments in the context of wireless sensor networks.

### C. Boids Model of Flocking Animals

Craig Reynolds [20] first presented a computer model of coordinated animal motion (such as bird flocks and fish schools) in 1987 called the "boids". The model is based on three separate but simple rules. The first rule is "separation" meaning steering to avoid crowding. Second rule is "alignment", which steers towards the average heading of local flockmates. And the third rule is "cohesion", which steers towards the average position of local flockmates. The "boids" model has been considered a model of realistic simulations ranging from SIGGRAPH presentations to Hollywood. Although the model has been extensively used, it has not been formally represented in the form of a framework. This need for a formal framework for an intelligent swarm has been pointed out by Hinchey et al. [26].

## III. A FORMAL FRAMEWORK

A formal framework [25] must comprise of three elements: first, it must provide for common concepts and terms precisely and unambiguously. Second, it must be sufficiently well-structured for future expansion and third, it must allow for alternate designs of models for the sake of comparison. In the following subsections, we first present a formal specification model of a WSN and subsequently, we develop a formal specification model of a self-organizing flocking "boids" system.

*A. A Formal Specification Model of WSN*

As first steps towards a formal description of a wireless sensor network, we need to be able to describe at least a certain number of key concepts:

```
_Location______________________
x:ℤ
y:ℤ
_______________________________
x≤max-x
x≥min-x
y≤max-y
y≥min-y
```

These include a model that defines the sensors themselves as well as a location which can be assigned to individual sensors as given above. Here we consider that the location consists of a point in a 2-D Cartesian coordinate system. Here, we see that the "Location" schema. After defining the Location, we need to initialize the values *max-x, min-x, max-y, min-y*. So, in the *InitLocation* schema, we are initializing the constants associated with the *Location*. These constants give the limits of the observed "world".

```
_InitLocation__________________
Location
_______________________________
max-x =35
min-x =-35
max-y =35
min-y =-35
```

Next we define a Sensor schema. Since the goal of this WSN is to monitor flocking "Boids", let us assume that the sensor is an advanced "cognitive" node which can monitor its surroundings and keep a count of "boids" in its immediate sensing radius or at the very least, be able to detect a boid in its proximity. This can be represented as following:

```
_Sensor________________________
Location
countNearbyBoids:ℕ
_______________________________
countNearbyBoids ≥0
```

Here we see that first of all, the *Sensor* schema includes the Location schema. Whereas the schema also includes a countable number "*countNearbyBoids*", which represents the sensed number of boids. In the predicate section of the schema, we represent the predicate saying that the *countNearbyBoids* will be a number greater than or equal to zero. This seems redundant because its data type is itself "fat" N, however the reason we have included it here is that "Z" specification is all about writing concepts formally and we do have a certain motive in saying that we expect nearby Boids to be greater than or equal to zero.

*B. Formal specification of the "Boids" model*

The formal specification of sensors was not that hard. However, the "Boids" model is not very simple because of a variety of reasons:
1. Firstly, in this model, there are an arbitrary number of seemingly randomly moving "boids".
2. These "boids" still flock together using nature-inspired rules.

So, the big question comes as to how to go about modeling this "Complex Adaptive System" with an emergent phenomenon of flocking using "Z". To solve just this problem, what we do here is that we first define the number of concepts that will be acceptable to clearly and completely define the concepts of the "Boids" world. "Z" allows fine tuning to the required level of abstraction by the use of something called the "given sets". In a given set, unlike a state schema, we do not have to go into the details of the system. Now, the emergent flocking behavior of the "boids" is based on certain phenomenon defined as following:

- "Alignment" requires the "boid" to have the tendency to turn so that it is moving in the same direction that nearby "boids" are moving.
- "Separation" means that a "boid" will also try to keep a certain distance away from other "boids" to avoid getting too close.
- "Cohesion" essentially makes the "boid" move towards other nearby birds unless it gets too close to another "boid".

Overall, when two birds are too close, the "separation" rule overrides the other two, which are deactivated until the minimum separation is achieved. The above three rules have an effect on the direction of motion only. Other than that, the "boids" keep moving forward at a "constant" speed. So, let us first start by defining an "axiomatic". This axiomatic is depicting that we have the given set BOID, which we define as the set of all boids. minSeparation is a value which effects the minimum separation between the "boids". maxAlignTurn is the maximum turn, a "boid" will do when turning for the sake of alignment of its heading with the average headings of its nearby flockmates. maxCohereTurn is the maximum turn that a "boid" will take when trying to get in coherent or in other words "closer" to its nearby flock mates. Similarly maxSeparateTurn is the maximum turn it will take to get separate from other flockmates if it ever gets too close. Vision is the radius of the boids, which it considers to select its neighbors. Although the rest are also angles, we selectively specify vision separately here since it is something which is more important in terms of the eventual flock size as compared to the other variables.

```
[BOID]
minSeparation:ℕ
maxAlignTurn:ℕ
maxCohereTurn:ℕ
maxSeparateTurn:ℕ
vision:ℕ
─────────────
vision ⩾ 0
vision ⩽ 360
```

Now, to model the boid, we need to further define a schema which gives attributes to it, defined as following:

```
─HeadedBoid──────────────
boid:BOID
loc: Location
heading:ℕ
─────────────
heading ⩽ 360
```

Here the *HeadedBoid* schema is showing that it contains a BOID along with a heading and a Location and that the heading ranges from zero to 360 degrees. Now, we need also to model the concept of distance.

```
─GetDistance──────────────
HeadedBoid
b1:HeadedBoid
distance!:ℕ
─────────────
distance! = √((b1.loc.x − loc.x)² + (b1.loc.y − loc.y)²)
```

In the above schema, we find the distance between the two boids using a simple geometric displacement calculation using Pythagoras theorem. After calculating the distance, we need a schema for finding the neighbors of a HeadedBoid:

```
─FindFlockmates──────────────
HeadedBoid
BOID
n, neighbors!:BOID
─────────────
n=∅
∀x:BOID
    if GetDistance(x) ⩽ vision
    n = n ∪ {x}
neighbors! = n
```

Thus, in the above *FindFlockmates* schema, we see that it checks the distance of all BOID members and if it is less than *vision*, then the member becomes a neighbor or flockmate.

Next, we write a schema for the *FindNearestNeighbor*, which can be used to find the nearest neighbor by using the GetDistance operation schema.

```
─FindNearestNeighbor──────────────
HeadedBoid
FindFlockmates
neighbors:𝔽BOID
nearest!:BOID
─────────────
neighbors = FindFlockmates
nearest! = x:neighbors
∀ x:neighbors
    if GetDistance(x) < GetDistance(nearest!)
    nearest! = x
```

Next, we need to define a schema for *Separate*. Here in this schema, we see that the HeadedBoid will turn upto a maximum of the maxSeparateTurn away from the heading of the nearest HeadedBoid.

```
─Separate──────────────
ΔHeadedBoid
nearestBoid?:HeadedBoid
tempHead:ℕ
─────────────
tempHead = getAwayTurn(nearestBoid?)
if tempHead ⩽ maxSeparateTurn
    then heading′ = heading + tempHead
else
    heading′ = heading + maxSeparateTurn
```

Here we see the schema for the operation *Align*. In case of the schema for *Align*, we notice that we find the turn towards the average heading of the flockmates. Then based on the maximum allowed alignment turn value of *maxAlignTurn*, the new heading is updated.

```
_Align______________________________
ΔHeadedBoid
avgHeading?:ℕ
tempHead:ℕ
____________________________________
tempHead=getTowardsTurn(avgHeading?)
if tempHead ⩽ maxAlignTurn
    then heading'= heading + tempHead
else
    heading'= heading + maxAlignTurn
```

Here, we provide a schema for the *Flock* operation. In this schema, we can see that the schema includes all the necessary operations and state schemas. The first priority is minimum separation and if it is not fulfilled, *Separate* is called. If not, then *Align* and *Cohere* are invoked respectively.

```
_Flock______________________________
HeadedBoid
FindNearestNeighbor
Separate
Align
Cohere
Distance
____________________________________
nearest=GetNearestNeighbor
if GetDistance(nearest) ⩽ minSeparation
then
    Separate(nearest)
else
    Align
    Cohere
```

In the getTowardsTurn schema, we basically find a turn by using a function which subtracts the headings geographically.

```
_getTowardsTurn_____________________
ΞHeadedBoid
boid?:HeadedBoid
turn!:ℕ
____________________________________
turn! = subtractHeading (boid?.heading, heading)
```

Similar to *getTowardsTurn* schema, we have the *getAwayTurn* schema which inverts the two headings.

```
_getAwayTurn________________________
ΞHeadedBoid
boid?:HeadedBoid
turn!:ℕ
____________________________________
turn! = subtractHeading (heading, boid?.heading)
```

IV. TRANSLATED AGENT-BASED MODEL

In this section, we present the translated Complex Adaptive System Simulation model based on the concepts translated from the formal specification model.

*A. Simulation Environment*

The simulation environment is built using an agent-based simulation tool NetLogo [27]. The model has been developed to allow for a translation of the formal specification model to the agent-based simulation environment programmed in the Logo programming language.

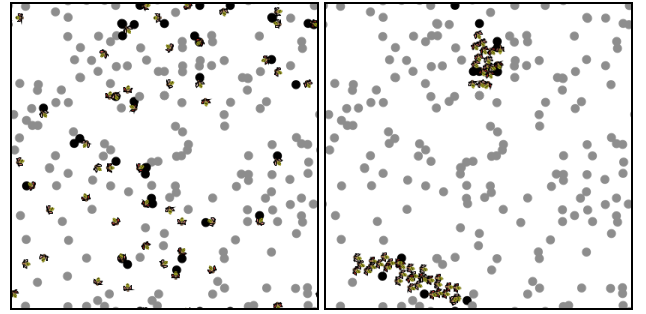

Figure 1(a) shows unflocked "boids" while (b) shows flocked "boids" and the sensor nodes sensing the boids, show up as black on the screen.

*B. Model Description*

Here we show the agent-based model built according to the formal specification given in previous sections. In Figure 1a, the sensors (shown as grey and black circles) and the boids, are shown as "bees". The sensors near the boids change color to black to represent proximity detection. In 1a, the boids are unflocked. After a certain amount of simulation time, the boids can be seen to flock together as shown in the model. The phenomenon of flocking can be seen in a more clear form in Figure 2. This is a time lapse picture showing only the tracks made by the boids. In Figure 2a, we see that the boids have just started and are trying to flock. In Figure 2b, however, we see that most of the boids have flocked together as evident by traces of their tracks.

V. DISCUSSION

In this section, we discuss the value-addition of the proposed framework for developing simulation models of wireless sensor networks in Complex Adaptive environments. In the absence of any comparative frameworks, here we examine how to design wireless sensor network simulations for monitoring Complex Adaptive environments in general. In particular, we also discuss how conventional modeling schemes fail to provide effective tools for building verifiable and credible simulation models for such complex environments.

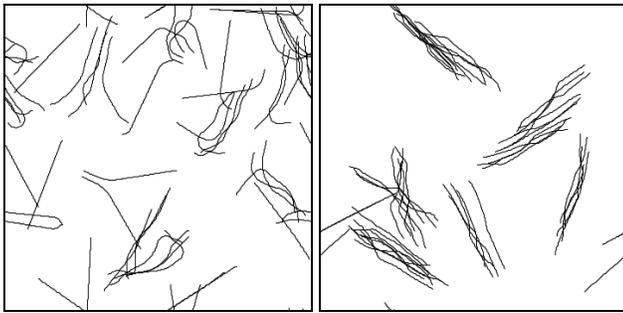

Figure 2 Sensed Motion tracks of "boids" showing emergence of flocking behavior (a) Unflocked (b) Flocked

### A. Problems in modeling Complex Adaptive Systems

Modeling requires an understanding of the system. Being able to develop a model demonstrates our prowess over a system. The "fidelity" of the developed model is representative of the level of confidence in comprehension of the original system. However, by definition, a "model" is a different concept from a "simulator" as described in [28].

In simple man-made systems, the assumption of having a model based on a complete understanding is possible. However, in the case of natural systems or even large scale artificial systems (such as P2P file sharing systems or WSNs or online social networks etc.), developing a complete understanding can be very hard if not impossible. Complex Adaptive Systems are known to exhibit collective behaviors such as emergence, self-organization and other similar phenomena. And if we couple a Complex Adaptive System (CAS) with Sensor Networks, which are already known to exhibit self-organization as discussed in [5], it becomes obvious that model building for sensing in Complex Adaptive Environments is a non-trivial task.

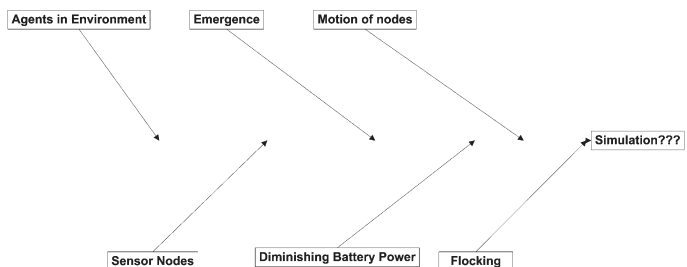

Figure 3 Fishbone diagram showing the sources of Complexity

The Cause-Effect (Fishbone) diagram illustrated in Figure 3 summarizes the sources of complexity in developing such models and simulators. Here, we see that problems such as having a large number of agents, randomly placed sensor nodes, emergence in the behavior of sensed agents, diminishing battery power as well as the mobility of nodes all jointly increase the level of complexity entailed in the development of simulation models.

### B. Possible Modeling approaches

Now, in the absence of a formal framework, there are two basic ways of developing computer simulation models of such hybrid complex adaptive systems. One way is to develop a traditional network model using a typical Network Simulator Tool (NS2/NS3/OPNET etc.) to develop the simulation. The other is to develop the simulation using an agent-based modeling approach[17]. Figure 4 summarizes these possibilities.

*1) Network modeling approach*

Firstly, as can be seen in the figure 4, the system can be modeled as a Network Model. This choice would however, lead to a problem in validation since the sensed values need to be chosen arbitrarily, perhaps in the form of some probabilistic distribution[29, 30]. Although probabilistic distributions are prevalent in literature, they can be used to represent simple phenomena such as arrivals, incidents of events etc. However, it is not possible to depict all aspects of behavior using only probability distributions. Because of the complexity of the environment, a probabilistic distribution based random value cannot always be considered a valid simplification of the actual environment.

An alternative approach to modeling the environment could be based on using optimization in general [31] or Particle Swarm Optimization (PSO) in particular. For example, Mohtasham et al. [32] have recently used a reactive monitoring mechanism termed "Simple Rate" in order to develop a minimal communication cost mechanism. The authors employed PSO to set local thresholds. The key difference with current work is in the use of a PSO based model versus our agent-based approach. Whilst there are also certain similarities apparent in a PSO based model since the optimization can be considered an agent based model (in the form of optimization of individual solutions), there are certain marked differences with the boids model. So, in the case of Complex Adaptive Environments such as boids, the agents need to interact with each other in order to form a realistic model of the self-organization mechanism related to flocking. Thus by definition, unlike a PSO based approach, realistic modeling of CAS would require each agent to interact with each other. In addition, every CAS type environment would require a modeling of different interactions amongst the agents in order to provide a realistic solution. Secondly, the model needs to be flexible since slight changes in parameters can result in an absence of flocking. This phenomenon can be seen e.g. when two different species of animals pass the same area. This can however also be modeled in a purely agent-based modeling paradigm such as FABS.

Furthermore, as Quera et al. [33] have noted, there is no simple index applicable to flocking in agent-based modeling that includes the range of factors describing the degree of flocking behavior. The authors also mention that flock-detection is normally carried out merely by observing the changes in the agents' locations over time on the computer screen. An interesting future possibility could be to combine

Mohtasham et al's PSO approach with our agent-based (FABS) framework in either of two ways:
1. Each agent could internally use PSO to optimize its alignment turns based on global alignment turns.
2. A formal approach could be used to develop a specification for an agent-based model combined with PSO as an extension to the FABS framework.

*2) Agent-based modeling approach*

As depicted in figure 4, we see that without the use of the proposed formal framework, agent-based modeling would require iterations of brainstorming and arbitrary definition of agents. This process is not a formal approach since it does not guarantee modeling of all desired agents and concepts. Besides, it also fails to ensure that we do not model more concepts as agents than are actually needed to solve our problem. While modeling insufficient concepts can result in an incomplete model, simulation of extraneous concepts can result in unnecessary complexity, errors and artifacts as discussed in [34]. In other words, without a formal specification mechanism, there is no way of confirming that the model is complete in all respects. Eventually this can result in a complex model lacking credibility [35].

With the proposed novel formal framework, we marry the benefits of using a powerful but easy to use formal specification language with the ease of developing agent-based models of Complex Adaptive Systems. Thus, we get the best of both worlds. It is quite well-known in literature that formal specification of models leads to completeness [21]. In other words, the formal specification is a real representation of the model in the desired "Experimental Frame" as given in [28]. Another interesting aspect about "Z" is that it allows for having an abstract description without details in the form of "given sets" as well as the capability to give detailed descriptions in the form of "schemas".

Subsequently, verification and validation exercises based on the developed formal specification can provide actual proofs of correlation with the real world. It would be pertinent here to mention that an appropriate methodology for Verification and Validation of Agent-based models has been recently proposed by the authors in [36, 37].

## VI. RELATED WORK

Current sensor network simulation toolkits are typically based on simulators such as NS-2, OPNET[38], J-Sim[39], TOSSIM[40] etc. Ballarini and Miller [41] describe verification of S-MAC, a medium access control protocol designed for wireless sensor networks, by means of the so called PRISM model checker however they do not propose any formal framework for formal specification of wireless sensor networks. In [42], a formal model, performance evaluation and model checking of WSNs is developed. The paper shows how a rewriting-logic-based Real-Time Maude language and tool can be used to formally model, simulate, and model check WSN algorithms. This paper focuses on the algorithms and uses Monte Carlo simulations. Another paper [43] describes KleeNet, a debugging environment that discovers bugs before deployment. KleeNet executes unmodified sensor network applications on symbolic input and automatically injects non-deterministic failures. SensorSim is a simulation framework [44]. Amongst traditional network simulations, J-Sim and NS2 have been compared in [39] and J-Sim has been shown to be more scalable. Atemu [45] on the other hand, focuses on simulation of particular sensors. Likewise, TOSSIM [40] is a TinyOS Simulator.

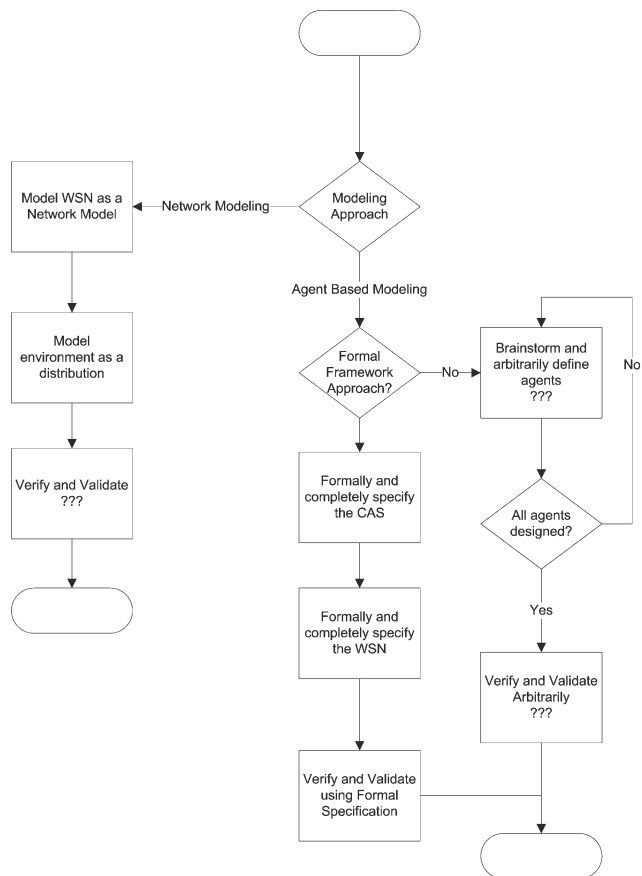

Figure 4 Modeling and Simulation of a WSN and a CAS

A global tracking system for animals has previously been suggested in [46] however it suggests using radio signals only and not Wireless Sensor Networks for this purpose. Other similar work is reported in [47] where tracking was performed for dragonfly migration by using attached transmitters followed using a Cessna aircraft. Heidemann et al. [48] discuss tracking animals using a WSN.

## VII. CONCLUSIONS AND FUTURE WORK

In this paper, we have presented FABS, a novel Formal framework for Agent-Based Simulation. We have used FABS for developing a simulation model for wireless sensor networks for monitoring Complex Adaptive Environments. We have demonstrated the usefulness of this novel methodology using a boids model of self-organized flocking of animals monitored by a random deployment of proximity sensors. Our preliminary work has demonstrated how formal specification can be effectively used to develop models of the Complex

Adaptive System Phenomena alongside the design of the wireless sensor network. The formal mathematical language employed, Z, allows for a clear, concise and unambiguous representation of the environment as well as the sensors and the sensor network. This framework can be easily extended to encompass other problems involving sensing of distributed entities and large scale areas to be monitored. In the future, we plan to demonstrate the application of this framework in sensing of other types of Complex Adaptive Environments such as sensors for sensing traffic flow and traffic jams.